\documentclass[12pt]{article}
\usepackage{amssymb,amsmath}
\begin{document}

\begin{titlepage}

                            \begin{center}
                            \vspace*{5mm}
\Large\bf{Vanishing largest Lyapunov exponent and Tsallis entropy}\\

                            \vspace{2cm}

              \normalsize\sf    NIKOS \  \ KALOGEROPOULOS $^\dagger$\\

                            \vspace{0.2cm}
                            
 \normalsize\sf Weill Cornell Medical College in Qatar\\
Education City,  P.O.  Box 24144\\
 Doha, Qatar\\

                            \end{center}

                            \vspace{2cm}

                     \centerline{\normalsize\bf Abstract}
                     
                           \vspace{3mm}
                     
\normalsize\rm\setlength{\baselineskip}{18pt} 

\noindent We present a geometric argument that  explains why some systems having vanishing largest 
Lyapunov exponent have underlying dynamics aspects of which can be effectively described by the Tsallis entropy. 
We rely on a comparison of the generalised additivity of the Tsallis entropy versus the ordinary additivity of the BGS entropy.
We translate this comparison in metric terms by using an effective hyperbolic metric on the configuration/phase space for the Tsallis 
entropy versus the Euclidean one in the case of the BGS entropy. Solving the Jacobi equation for such hyperbolic metrics
effectively sets the largest Lyapunov exponent computed with respect to the corresponding Euclidean metric to zero.  
This conclusion is in agreement with all currently known results about systems that have a simple 
asymptotic behaviour and are described by the Tsallis entropy.\\

                             \vfill

\noindent\sf  PACS: \  \  \  \  \  \  02.10.Hh, \  05.45.Df, \  64.60.al  \\
\noindent\sf Keywords: \ Tsallis entropy, \ Lyapunov exponents, \  Nonextensive statistical mechanics, \ CAT(k).  \\
                             
                             \vfill

\noindent\rule{8cm}{0.2mm}\\
\noindent \small\rm $^\dagger$  E-mail: \ \  \small\rm nik2011@qatar-med.cornell.edu\\

\end{titlepage}


                            \newpage

\normalsize\rm\setlength{\baselineskip}{18pt}

          \centerline{\large\sc 1. \ Introduction}

                                     \vspace{3mm}

The Tsallis entropy is a single-parameter family of functionals, first introduced in the Physics literature in 1988 [1], providing 
an alternative to the Boltzmann/Gibbs/Shannon (BGS) entropy used in the statistical description of a system.  
Consider a probability distribution  $\{ p_i \} , \ i\in I $ 
in a discrete sample space indexed by \ $I\subset \mathbb{N}$. \ Its Tsallis entropy is defined as
\begin{equation} 
        S_q = k_B \  \frac{1}{q-1} \left( 1 - \sum_{i\in I} \ p_i^q \right)
\end{equation}        
It may be worth comparing (1) with the BGS entropy
\begin{equation}
        S_{BGS} ( \{ p_i \} ) = - k_B \sum_{i\in I} p_i \log p_i
\end{equation}
where we immediately observe as the non-extensive/entropic parameter \ $q\rightarrow 1$, \ we get
\begin{equation}
        \lim_{q\rightarrow 1} S_q = S_{BGS}
\end{equation}
 An analogous definition of the Tsallis entropy can be given for continuous sample spaces. 
 Henceforth, we will be setting the Boltzmann constant \ $k_B = 1$, \  for simplicity.\\
 
 The Tsallis entropy, conjecturally, describes collective phenomena
with  long-range spatial and temporal correlations [2], [3], systems whose phase portraits exhibit a fractal-like behavior etc. 
for which there is no reason or justification why their description by the BGS entropy should be accurate or even 
valid [3] (and references therein). Following the approach and viewpoint of Boltzmann [4], [5], one can state that the 
dynamical basis of the Tsallis entropy remains unclear so far [3].  This mirrors the existing difficulties in deriving the BGS 
entropy [3]-[6] from dynamical principles. As in the latter case, in the case of the Tsallis entropy, some progress  
has been made in identifying characteristics of systems effectively described by it. One class of such systems are ones 
exhibiting ``weak chaos" or being at the ``edge of chaos" for some subset of their parameter space [2], [3], [7]-[10]. This is more 
accurately expressed by referring to them as dynamical systems having vanishing largest Lyapunov exponent [3], [7]-[10].\\

A general formal justification of why such systems are described by the Tsallis entropy 
has been lacking so far [3], [11]. We attempted to outline an explanation in our previous work [12]. 
In the present work we make things more concrete and also show how such results  are independent of and 
can be extended to the more general case of \ $\mathrm{CAT(k)}$ \  configuration/phase spaces. We would like 
point out some important predecessors of applications of (differential-) geometric methods in Statistics [13]-[19] 
and Probability [20], [21]. These references may be consulted 
for some relevant background as well as for providing a far broader and deeper perspective on the geometric methods 
employed in the present work, from the viewpoint of Statistics [13]-[19] and Probability [20], [21].\\ 
 
In Section 2, we argue that if a system is described 
by the Tsallis entropy and moreover if the Tsallis induced generalised addition is a direct result of its configuration/phase 
space dynamics, then the largest  Lyapunov exponent  of the underlying system should vanish. 
This statement is in agreement with all currently known numerical as well as the very few available analytical results [22]-[36]. 
In general, it is probably fair to say that few things are known about the analytical basis of such results, 
and this the issue that we partly intend to address in  the present work. 
A recently pointed out mismatch between numerical predictions [37]-[39] and analytical results [40] shows that we have to 
be careful when performing numerical extrapolations. On the other hand, [41] shows that both the numerical predictions and the 
analytical results can be in agreement with each other. 
In Section 3,  we extend this conclusion to interacting systems having different values of \ $q$. \
Such interacting systems should be  modelled on \ $\mathrm{CAT(k), \ k<0}$ \ spaces, as we pointed out in [42]. 
Section 4, contains some comments and points toward further implications of the Tsallis 
entropy composition property (6), employing the underlying concept of hyperbolicity.\\

For brevity, we do not provide the required background in Riemannian geometry  
or the metric geometry of \ $\mathrm{CAT(k), \ k<0}$ \ spaces, referring instead to some of the excellent 
references such as [43], [44] and [21], [45]  respectively, on these topics, in addition to the ones provided above.\\

                                   \vspace{5mm}


          \centerline{\large\sc 2. \  Riemannian spaces: vanishing largest Lyapunov exponent}

                                        \vspace{5mm}
 
One of the  features distinguishing the Tsallis  (1) from the BGS (2) entropy is their different composition properties. This distinction has
profound consequences for the definitions of independence and additivity/extensivity that pervade Statistical Mechanics and 
Thermodynamics [3] 
(and references therein). Two systems \ $A$ \ and \ $B$ \ are conventionally defined to be ``independent" if in their statistical description, the
corresponding probability distributions \ $p_A$, \ $p_B$ \ have as composition law the ordinary multiplication
\begin{equation}
   p_{A + B} = p_A \cdot p_B
\end{equation}
Here and henceforth \ $A+B$ \ indicates the compound system formed by combining \ $A$ \ and \ $B$. \
For such independent systems (4), the BGS entropy is additive
\begin{equation} 
        S_{BGS} (A+B) = S_{BGS}(A) + S_{BGS}(B) 
\end{equation} 
as can be immediately seen from (2). By contrast, under the same definition of independence (4),
the Tsallis entropy (1)  obeys the generalized additivity/composition property 
\begin{equation}
  S_q (A + B) = S_q (A) + S_q (B) + (1-q)  S_q(A) S_q(B) 
\end{equation}
This composition motivated the introduction of a generalized addition [46], [47] by
\begin{equation} 
       x\oplus_q y = x+ y + (1-q) xy 
\end{equation}  
The generalised addition (7) stemming form the Tsallis entropy composition property (6) is arguably the biggest difference between the 
BGS and the Tsallis entropies.  It is worth observing that
\begin{equation}
    x\oplus_q y \ \sim \ x + y, \hspace{10mm} |x| \ll1, \ |y| \ll1 
\end{equation}
and that asymptotically 
\begin{equation}
    x\oplus_q y \ \sim \ xy, \hspace{10mm}  |x| \rightarrow \infty, \ \  |y| \rightarrow \infty
\end{equation}
Therefore the difference between (7) and the ordinary addition should becomes significant in the asymptotic regime 
\ $|x|, |y| \rightarrow \infty$ \ describing very highly entropic systems, such as black holes, for instance. So, we expect that the differences 
between
the BGS and the Tsallis entropies express the different ways that they describe highly entropic systems. For such systems, the Tsallis entropy 
induces an essentially ``exponential way" of combining interacting systems as is evident from (9). 
This should be contrasted with the  ``linear"  fashion in which  the BGS entropy performs the same task.\\
 
It took sometime to find a generalised product  that would be distributive with respect to (7), so together they would form a nice algebraic 
structure expressing the non-trivial composition properties of the Tsallis entropy versus the corresponding properties of the BGS entropy.
Actually, two such generalized multiplications were introduced independently in [48], [49]. Although 
conjecturally independent, an explicit equivalence between them is still lacking. 
In these works a field isomorphism was, essentially, introduced
that was indicated by \ $\tau_q$ \ in [49] where some of its rudimentary metric and measure-theoretical properties on \ $\mathbb{R}$ \ 
were examined.  In our subsequent work [42], we developed further some of the metric consequences of \ $\tau_q, \ 0\leq q < 1$, \ 
initially for \ $\mathbb{R}^n$. \  Motivated by an analogy with the translation invariance of the Euclidean metric, 
 we constructed in [42] a Riemannian metric \ {\bf g}, \ induced by (6), with components 
\begin{equation}
     {\bf g} = \left( \begin{array}{ll}
                                1 &  0 \\ 
                                0 & e^{-2tx}
                            \end{array}\right)
\end{equation}
and corresponding line element 
\begin{equation}
    ds^2 = dx^2 + e^{-2tx} dy^2
\end{equation}
where
\begin{equation}
    t = \log (2-q)
\end{equation}
The BGS composition property (``ordinary addition")  is expressed, in this formalism, via the effective Euclidean metric  
\begin{equation}
      {\bf g}_E = \left( \begin{array}{ll}
                                1 &  0 \\ 
                                0 & 1
                            \end{array}\right)      
\end{equation}
whose corresponding line element is 
\begin{equation}
    ds^2_E = dx^2 + dy^2
\end{equation}
The two metrics (10), (13) were used in [42] to encode and compare, in metric terms, the composition properties of the 
BGS and the Tsallis entropies. \\

The metric tensor (10)  turned out [42] to have a constant negative sectional curvature 
\begin{equation}       
    k = - [ \log (2-q) ] ^2
\end{equation}
which provided a geometric interpretation of the non-extensive parameter \ $q \in [0, 1) $. \ 
Endowing the plane \ $\mathbb{R}^2$ \ with (7),  turned it into a re-scaled version of the hyperbolic plane \ $\mathbb{H}^2$. \  
Then, we concluded in [42], that due to (7) which gave rise to (10), the Tsallis entropy can be thought as a 
``hyperbolic analogue" of the BGS entropy.\\

The subsequent discussion will assume that the generalised addition (7) is somehow manifest at the level of configuration/phase 
space of the system. Although such spaces are Riemannian manifolds, therefore locally Euclidean, to understand the difference 
between the Tsallis and the BGS entropies, we should be actually looking at their large-scale/ large-distance, asymptotic properties [45], 
[21], [50].  This is suggested by a comparison between (9) and the ordinary addition. In a sense it is the inverse of 
what Statistical Mechanics attempts to do: here we use the thermodynamic additivity to determine the effective concept of additivity that is 
applicable in the configuration/phase space of the underlying dynamical system.  
Naturally the present assumption is not valid if the additivity properties of the thermodynamic quantities are ``emergent" 
in some non-trivial way from the underlying dynamics.\\   

To continue and without actually  adding any essential complexity to the subsequent argument, we will consider in the formalism instead 
of just \ $\mathbb{R}^2$,  \ a general Riemannian manifold \ ($M, {\bf g}$) \ with tangent bundle indicated by \ $TM$. \ This \ $M$ \ could 
represent  the configuration or the phase space of a physical system as noticed in the previous paragraph.  Let the Levi-Civita connection 
compatible with 
\ ${\bf g}$ \  be indicated by \ $\nabla$. \ Such a connection is expressed in terms of \  ${\bf g}$ \ 
by the Koszul formula
\begin{eqnarray}  
  2 {\bf g} (\nabla_XY, Z) = & - {\bf g}(X,[Y,Z]) + {\bf g}(Z,[X,Y]) + {\bf g}(Y,[Z,X]) \nonumber \\
                                                &     + X[{\bf g}(Y,Z)] - Z[{\bf g}(X,Y)] + Y[{\bf g}(X,Z)]                                                   
\end{eqnarray}
where \ $X, Y, Z \in TM$. \ 
The geodesic equation is [43] - [45]
\begin{equation} 
          \nabla_X X = 0
\end{equation}
for \ $X \in TM$ \ tangent to the geodesic. If \ $J \in TM$ \ indicates a Jacobi field then it satisfies the Jacobi/geodesic deviation 
equation [43], [44] 
\begin{equation}
   \nabla_X\nabla_X J + R(J, X)X = 0
\end{equation}
Here \ $R(X, Y)Z $ \ indicates the Riemann tensor which is defined by [43], [44]
\begin{equation}
   R(X,Y)Z = \nabla_X\nabla_Y Z - \nabla_Y\nabla_X Z - \nabla_{[X,Y]} Z
\end{equation}
It is extremely difficult to explicitly solve either the 
geodesic (17) or the Jacobi equations (18), except in a few particularly simple cases, our case of interest being one of them. 
Let \ $e_1, e_2$ \ be orthonormal vectors, with respect to \ ${\bf g}$, \ spanning a 2-dimensional subspace of \ $TM$ \ in a 
neighborhood of \ $x \in M$. \ The sectional curvature of \ $M$ \ in this 2-plane subspace of \ $TM$ \ is defined by [43], [44]
\begin{equation}
    k = {\bf g}(R(e_1, e_2)e_2, e_1) 
\end{equation}
In our case of interest, initially at  least, \ $k < 0$ \ is constant.  
Then the Jacobi equation (18)  reduces to the  ordinary differential equation 
\begin{equation}
    \frac{d^2 J(s)}{ds^2} - k J(s) = 0  
\end{equation}
which has the general solution       
\begin{equation}
    J(s) = \sum_{i=1} ^{n-1} \left\{ a_i \exp (\sqrt{-k} \ s) + b_i \exp  (-\sqrt{-k} \ s) \right\} \ e_i(s) 
\end{equation}
where \ $a_i, \ b_i$ \ are constants, \ $\{ e_i(s) \} $ \ are parallel orthonormal vectors (Fermi basis) \ and \ $s$ \ is the arc-length parameter of 
the geodesic whose  tangent is \ $X(s)$. \ The summation takes place over the \ $n-1$ \ directions orthogonal to \ $X(s)$. \  
Substituting (15) into (22), we  find that
\begin{equation} 
    J(s) = \sum_{i=1} ^{n-1} \{ a_i (2-q) e^s - b_i  (2-q) e^s \} \ e_i(s)     
\end{equation}
Hence, we see that in a Riemannian manifold of constant negative curvature (15), the nearby geodesics deviate from each other 
exponentially in terms of the geodesic arc-length \ $s$ \ or any of its affine re-parametrizations. The Riemannian metric (10) is a 
special case of the general \ ${\bf g}$, \ which is defined on \ $M = \mathbb{R}^2$. \  This exponential deviation of the 
nearby geodesics of a manifold of negative sectional curvature should be contrasted to those of the Euclidean metric (13) which has \ $k=0$: 
in the Euclidean case the geodesics separate linearly 
as functions of the arc-length parameter \ $s$, \ as can be immediately seen  from (22) by setting \ $k=0$. \\

 Let's rephrase the above argument in an alternative, coordinate-dependent, way.  Consider a unit vector with respect to (13). 
 This vector will have a smaller magnitude with respect  to (10), as can be 
immediately seen.  So (10) does not increase the magnitudes of the unit vectors. Instead, it exponentially decreases their $y$-component.
As a result,  it will  increase distances of any rectifiable curve in the $y$/transversal direction to $x$, by an exponential factor, 
which is exactly the statement contained in (22).\\ 

We turn our attention to Lyapunov exponents. We will only be needing their definition for the case of uniformly hyperbolic dynamical 
systems [51], [52], a standard example of which is the geodesic flow on a Riemannian manifold of, generally variable, 
negative sectional curvature. To be slightly more general than that, let \ $f_t: M \rightarrow M$ \ be a flow on the Riemannian 
manifold \ ($M, {\bf g}$), \ whose generating vector field  is    
\begin{equation}
   X(t) = \frac{d}{dt}(f_t(x))|_{t=0}
\end{equation}
Let  \ $Y \in TM$ \ and let its norm with respect to the Riemannian metric \ ${\bf g}$ \ be indicated by \ $ ||Y|| = \{ {\bf g}(Y,Y) \}^{\frac{1}{2}} $. 
\  The Lyapunov exponent of  the perturbation in the direction of \ $Y\in T_xM$ \ along the evolution/trajectory 
of the flow  \ $f_t$ \ is defined by
 \begin{equation}    
      \lambda_x(Y) = \lim_{t\rightarrow\infty} \frac{1}{t} \log ||d_x(f_tY)||
\end{equation}
The Lyapunov exponent measures the asymptotic rate of change of the magnitude of a perturbation in the direction of \ $Y$. \
Is is clear that if someone is interested in a stability analysis of a flow, the most pertinent Lyapunov exponent  
is the largest positive one in some transversal direction. Consider, as a special case, \ $f_t$ \ to be the geodesic flow on \ $\mathbb{R}^2$ \ 
endowed with (8) that was discussed above. We see from (23) that  
\begin{equation}
   \lim_{s\rightarrow \infty}  ||J(s) || = e^s
\end{equation}
therefore, the definition (25) gives
\begin{equation} 
     \lambda = 1
\end{equation} 
To summarize: we started from a dynamical system modelled on a Riemannian manifold \ $M$ \ equipped with a metric induced by (14). 
We assumed that this dynamical system's effective statistical description is provided by the Tsallis entropy (1). 
In turn, the Tsallis entropy composition property (6) indicated that it is more suitable to use the effective hyperbolic metric (10) on \ $M$ \
manifold, instead of the original one (13). It was the collective description of the underlying dynamics 
that dictated this ``hyperbolization" in producing the effective metric (10) from (13). 
This ``hyperbolization" was concretely implemented as the ``warping" by a convex function, 
the exponential in the present case, in transitioning from (13) to (10). The instabilities of the dynamical system were expressed through its 
positive Lyapunov exponents, initially with respect to (13), the largest of them being the most important. 
 When the effective behavior  described by the hyperbolic metric (10) is taken into account, the  largest positive Lyapunov exponent of the 
underlying dynamical system becomes zero, as can be seen in (26) and (27). 
The reason is that the perturbations of the underlying dynamical system and the distances with respect to 
the hyperbolic metric (10) grow at the same rate as seen in (26). Hence their relative growth rate is linear, or could more generally be 
polynomial/power-law,
as seen in (27), so the Lyapunov exponent of the underlying dynamical system relative to (10)  is zero.  
This is the main conclusion of the present work and it is in agreement with all known 
results reached by analyzing particular models  [3] and references therein, as well as [22]-[36] for some, mostly numerical, 
results in this direction. \\   

In case the largest Lyapunov exponent is zero, as in the case discussed above, the instabilities of the dynamical system grow at a milder 
than exponential rate with respect to the hyperbolic metric (10), and consequently their asymptotic behavior  has to be encoded differently,
if we want to obtain non-trivial results. One way to quantify the growth of such perturbations  by modifying the definition of Lyapunov 
exponents (28) to    
 \begin{equation}
         \tilde{\lambda}_x(Y) = \lim_{t\rightarrow\infty} \frac{ \log ||d_xf_tY||}{\log t}
 \end{equation}
These modified exponents describe perturbations obeying an asymptotic, power-law evolution \ $t^{\tilde{\lambda}_x(Y)}$. \  
This is essentially the definition adopted in [3], [7]-[9]
\begin{equation} 
   \frac{d\xi}{dt} = \lambda_q \xi^q  
\end{equation}
for systems described by the Tsallis entropy, in just different notation. Naturally, the modified Lyapunov exponents (29) will explicitly 
depend on the value of the entropic parameter \ $q$ \ of the dynamical system.  Since, as is conjectured, the systems described by 
the Tsallis entropy may possess more than just one value of non-extensive parameter $q$ \ [3], [11], depending on which property 
of  the system is described by them, the modified Lyapunov exponents (29) would evidently depend on the value of 
\  $q$ \ that determines the sensitivity  of the system to infinitesimal perturbations, indicated as \ $q_{\mathrm{sen}}$ \  in [3].\\

The use of a definition like (29) would be of limited interest for applications, if explicit constructions/examples did not exist for which it 
could provide non-trivial information. 
Motivated by the linear (exponential) increase of the geodesic distances in Riemannian manifolds of zero (negative resp.) sectional 
curvature (23), one may be wondering whether an intermediate behavior such as described by (29) is even possible. The answer turns out 
to be affirmative, but it is not possible in the context of Riemannian manifolds. 
An example of a quadratic separation of geodesics in a 2-complex endowed with a CAT(0) metric was 
constructed \footnote{We are grateful to Professor Panos Papasoglou for bringing this work to our attention.}
in [53]. 
This quadratic geodesic deviation in a CAT(0) space is quite different from the case of \ $\mathrm{CAT(k), \ k<0}$, \ whose geodesics 
deviate exponentially from each other, as will be explained in the next Section. 
Whether the property of quadratic or, more generally,  polynomial divergence of geodesics encoded in (29) is ``typical", or even common, 
for CAT(0) spaces, and whether in such a case the definition (29) exhausts all possible geodesic deviation behavior, 
does not seem to be known at this time.  More importantly, it is unclear, to us at least, exactly what, if any, physical system the CAT(0) space 
construction of [53] can be used to describe. For these reasons, we refer to [53] for the construction itself and its mathematical aspects, 
aiming to re-visit this topic, if  physical reasons warrant it in the future. \\

                                                                          \vspace{5mm}


   \centerline{\large\sc 3. \  \ Geodesic \ deviation \ in \  \ $\mathrm{CAT(k), \ k<0}$ \ \ spaces}

                                                                     \vspace{5mm}
 
We continue with the generalization of the above results to the case of \ $\mathrm{CAT(k), \ k<0}$ \ spaces. 
The need for using \ $\mathrm{CAT(k), \ k<0}$ \ spaces, motivated by the composition of the Tsallis entropy (6),  was explained in [12], [42]. 
The major obstacle in repeating the Riemannian approach verbatim, is that  \ $\mathrm{CAT(k)}$ \  spaces do not possess a differential 
structure [21], [43]-[45] so one has to dispense with statements relying on regularity properties, such as ones formulated via vector fields, 
the geodesic (17) and the Jacobi (18) equations etc. The only option left is to employ the triangle inequality, which when combined with the 
\ $\mathrm{CAT(k), \ k<0}$ \ condition proves to be sufficient for attaining the sought-after goal [21], [45], [50]. One should notice, 
that due to lack of smoothness, most statements in the present Section can only be formulated via inequalities, as contrasted to the 
equalities,  such as (22), derived in the Riemannian case. As a result, the arguments in this Section, applied to  
the case of \ $\mathrm{CAT(k), \ k<0}$ \ spaces are, inevitably, synthetic as opposed to the analytic ones in the Riemannian case of the 
previous section. We will, largely, follow [50] in the sequel.\\    

One begins by  realizing that a geodesic space \ $(\mathfrak{X},d)$ \ which is \ $\mathrm{CAT(k), \ k<0}$ \ is necessarily hyperbolic. 
There are several definitions of hyperbolicity at various levels of generality [50], [45] and several equivalences among them [50], [45]. 
The following definition, ascribed to E. Rips, is the most useful for our purposes:  a metric space \ $(\mathfrak{X}, d)$ \ is   
\ $\delta$-hyperbolic, for \ $\delta >0$, \ if for any three points 
\ $x, y, z\in \mathfrak{X}$, \  any side of the triangle having \ $x, y, z$ \ as vertices lies in a \ $\delta$-neighborhood of the union of the two 
others.  Given this definition, the hyperbolicity of  the \ $\mathrm{CAT(k), \ k<0}$ \ space \ $(\mathfrak{X}, d)$ \ follows immediately from the 
definition of  the \ $\mathrm{CAT(k)}$ \ condition.\\

Consider a polygon in such \ $(\mathfrak{X}, d)$ \  with \ $n\in\mathbb{N}$ \ vertices \ $x_1, x_2, \ldots x_n$. \  Let 
\begin{equation}
            m = \left[ \frac{n}{2} \right] +1
\end{equation}
where the square brackets indicate the integer part of their argument. Because \ $(\mathfrak{X}, d)$ \ is a geodesic space, every segment has 
a midpoint. Consider the midpoint \ $x_m$ \ and the triangle \ $x_1 x_m x_n$. \ Since \ $(\mathfrak{X}, d)$ \ is \ $\delta$-hyperbolic, there is a
point \ $y$ \ in the union of the two other sides \ $x_1 x_m \cup x_m x_n$ \ such that \ $d(y, z) \leq \delta$ \ where \ $z\in x_n x_1$. \ 
Without loss of generality, we assume that \ $y\in x_1 x_m$. \  Using induction on the $m$-polygon \ $x_1, x_2, \ldots, x_m$, \ 
we find that the distance of \ $z$ \ from  the union of all other sides of this $m$-polygon is
\begin{equation}   
     d \left( z, \ \bigcup_{i=1}^{m-1} x_i x_{i+1} \right) \leq (p-1)\delta
\end{equation}
where \ $p\in \mathbb{N}$ \ satisfies 
\begin{equation}
                      p \geq \frac{\log (n-1)}{\log 2}
\end{equation}
We conclude then, that in \ $(\mathfrak{X}, d)$ \ each side of the $n$-polygon is contained in a $p\delta$-neighborhood of the union of its 
other sides. \\     

The second, and last, step is the quantification of the concept of the exponential separation of geodesics in \ $(\mathfrak{X}, d)$. \
Consider two segments emanating from \ $x\in\mathfrak{X}$ \ toward \ $y, \ z \in \mathfrak{X}$, \ respectively. 
Let two objects move with unit speed, one in each of these two arc-length parametrized segments. 
We are interested in the separation of \ $x$ \ and \ $y$ \ after time \ $t$. \ An obvious way to measure such a separation would be to 
start at the location of \ $y$, \ move back along the segment joining it with \ $x$ \ by a distance \ $t$, \ switch segment at \ $x$ \ and then 
continue from \ $x$ \ alongside the other segment for an additional distance \ $t$ \ until reaching \ $z$. \ This is not however, the analogue of a 
curve  joining \ $y$ \ and \ $z$ \ of Section 2. What we want is to measure the ``direct" separation between \ $y$ \ and \ $z$, \ without having to 
go back close to the intersection \ $x$ \  of the two segments.  
So, we want to determine the length of a path from \ $y$ \ to \ $z$, \ in the complement of a ball \ $B_r(x)$ \ of radius \ $r>0$ \ 
centered at \ $x$. \ Consider a path \ $x_1 x_2 \ldots x_n, \ n \in\mathbb{N}$, \ such that \ $d(x_i, x_{i+1}) \leq \epsilon, \ \  i=1,\ldots n-1$ \ 
which lies outside the ball \ $B_r(x)$, \ with \ $x\in x_n x_1$. \ Let \ $p\in\mathbb{N}$ \ be as in (30). \ Using the conclusion of the previous 
paragraph, there exists a point \ $w \in x_i x_{i+1}, \ i=1, \ldots, n-1$ \ such that \ $d(x, w) \leq p\delta$. \ Since \ 
\begin{equation}
d(x, w) \geq r - \frac{\epsilon}{2}
\end{equation} 
we find 
\begin{equation}
 p \geq \frac{r}{\delta} - \frac{\epsilon}{2\delta}
\end{equation}
which gives 
\begin{equation}
n \geq 2^{p-1} \geq c\cdot 2^{\frac{r}{\delta}}
\end{equation}  
where
\begin{equation}
c = 2^{-(\frac{\epsilon}{2\delta} +1)}
\end{equation} 
The conclusion that we reach from (35), is that the number of points with a uniform distance upper bound \ $\epsilon$ \ between two 
consecutive ones, making up the path between \ $y$ \ and \ $z$ \ outside \ $B_r(x)$ \ increases exponentially with \ $r$ \ in the 
\ $\mathrm{CAT(k), \ k<0}$ \ space \ $(\mathfrak{X}, d)$. \ This is what we wanted to show. We see that (35) is the analogue of (23) 
for \ $(\mathfrak{X}, d)$. \ Naturally, the argument leading to (35) is already applicable 
to the case of a Riemannian manifold \ $(M, {\bf g})$ \ of negative sectional curvature, since in this case \ $(M, {\bf g})$ \ is a
 \ $\mathrm{CAT(k), \ k<0}$ \ space. So, the conclusions drawn in the Riemannian case, about the effective metric behavior 
 of systems described by the Tsallis entropy, can be extended unaltered to the case of any number of interacting systems 
 described by different values of \ $q$.\\    
   
                                                                          \vspace{8mm}


   \centerline{\large\sc 4. \  \  Conclusions \ and \ Outlook}

                                                                     \vspace{5mm}
        
In the present work, we attempted to justify why dynamical systems whose statistical behavior is described by the Tsallis entropy, have
vanishing largest Lyapunov exponent. This was essentially ascribed to employing the effective negative curvature metric (10), which is  the 
``hyperbolization" of the Euclidean initially employed metric (13), as was pointed out in [12], [42]. Moreover we made the very strong 
assumption that the  additivity properties of the configuration/phase space of the system are directly reflected on its thermodynamic additivity  
(7) which is not emergent in any non-trivial manner. Our conclusion is in agreement with all currently known results.\\
 
The process of generalizing this conclusion to the case of \ $\mathrm{CAT(k), \ k<0}$ \ spaces     
presented in Section 3, is of interest, as it points out to the underlying reason behind such behavior.  
The argument of Section 3 shows that the key in understanding consequences of the Tsallis entropy composition property (7) 
is the concept of hyperbolicity, which was also alluded to in [12], [42]. Knowing this, is should not come as a surprise that the argument of
Section 3 is a small part of a well-known proof of Morse's Lemma, which establishes  the stability of geodesics in hyperbolic geodesic 
spaces under quasi-isometries [21], [45], [50]. In particular, since Riemannian manifolds of negative sectional 
curvature \ $(M, {\bf g})$ \ are a subset of \ $\mathrm{CAT(k), \ k<0}$ \ spaces, the vanishing of the highest Lyapunov exponent of the 
dynamical systems modelled by \ $(M, {\bf g})$ \ just expresses their underlying hyperbolicity in a very compact way. 
The general framework of this hyperbolicity and its implications for systems described by the 
Tsallis entropy will be examined further in a future work.\\

                                                                          \vspace{5mm}
 
 
                                                \centerline{\large\sc Acknowledgement}  

                                                                           \vspace{5mm}

The author is grateful to Professor Constantino Tsallis for emphasizing that it is the largest Lyapunov exponent that 
vanishes in systems described by the Tsallis entropy, but not necessarily all of them. The author is also grateful to the 
anonymous referee for pointing out to him the work of Amari and collaborators, as explained in [16], [17], [19],  
that bears some commonalities with the methods employed in the present work.\\

                                                                           \vspace{8mm}


                                                        \centerline{\large\sc References}

                                                                       \vspace{5mm}
                               
\noindent [1] \  C. Tsallis, \  \emph{J. Stat. Phys.} {\bf 52}, \ 479 \  (1988)  \\
\noindent [2] \  A.M. Mariz, C. Tsallis, \ \emph{Long memory constitutes a unified mesoscopic mechanism\\  
                           \hspace*{6mm} consistent with nonextensive statistical mechanics}, \ arXiv:1106.3100\\
\noindent [3] \  C. Tsallis, \ \emph{Introduction to Nonextensive Statistical Mechanics: Approaching  \\
                           \hspace*{6mm}   a Complex World}, \ Springer \ (2009)\\
\noindent [4] \ L. Boltzmann, \  \emph{Acad. Wissen. Wien, Math.-Naturwissen.} {\bf 75}, \  67 \ (1877)\\
\noindent [5] \ G. Gallavotti, \ \emph{Statistical Mechanics: A Short Treatise}, \ Springer \ (1999)\\
\noindent [6] \ E.G.D. Cohen, \ \emph{Pramana} {\bf 64}, \ 635 \ (2005)\\
\noindent [7] \ P. Grassberger, M. Scheunert, \ \emph{J. Stat. Phys.} {\bf 26}, \ 697 \ (1981)\\
\noindent [8] \ G. Anania, A. Politi, \ \emph{Europhys. Lett.} {\bf 7}, \ 119 \ (1988)\\
\noindent [9] \ H. Hata, T. Horita, H. Mori,  \ \emph{Prog. Theor. Phys.} {\bf 82}, \ 897 \ (1989)\\
\noindent [10] \ M.A. Fuentes, Y. Sato, C. Tsallis,  \ \emph{Phys. Lett. A}  {\bf 375}, \ 2988 \ (2011) \\
\noindent [11] \ C. Tsallis, \ \emph{Some Open Points In Nonextensive Statistical Mechanics}, \ arXiv:1102.2408\\
\noindent [12] \ N. Kalogeropoulos, \ \emph{QScience Connect}, \ 2012:12\\
\noindent [13] \ C.R. Rao, \ \emph{Bull. Calcutta Math. Soc.} {\bf 37}, \ 81 \ (1945)\\  
\noindent [14] \ I. Csisz\'{a}r, \ \emph{Ann. Prob.} {\bf 3}, \ 146 \ (1975)\\
\noindent [15] \ B. Efron, \ \emph{Ann. Stat.} {\bf 3}, \ 1189 \ (1975)\\
\noindent [16] \ S. Amari, \ \emph{Ann. Stat.} {\bf 10}, \ 357 \ (1982)\\
\noindent [17] \ S. Amari, \ \emph{Differential-Geometrical Methods in Statistics}, \ Springer (1985)\\ 
\noindent [18] \ O.E. Barndorff-Nielsen, \ \emph{Ind. J. Math.} {\bf 29}, \ 335 \ (1987)\\
\noindent [19] \ S. Amari, H. Nagaoka, \emph{Methods of Information Geometry}, Amer. Math. Soc. (2000)\\  
\noindent [20] \ S.T. Rachev, \ \emph{Probability Metrics and the Stability of Stochastic Models},  Wiley (1991)\\
\noindent [21] \  M. Gromov, \ \emph{Metric Structures for Riemannian and Non-Riemannian Spaces},\\
                     \hspace*{8mm}  Birkh\"{a}user \ (1999)\\
\noindent [22] \ C. Tsallis, A.R. Plastino, W.-M. Zheng, \ \emph{Chaos, Sol. and Fractals} {\bf 8}, \ 885 \ (1997)\\
\noindent [23] \ U.M.S. Costa, M.L. Lyra, A.R. Plastino, C. Tsallis, \ \emph{Phys. Rev. E} {\bf 56}, \ 245 \ (1997)\\
\noindent [24] \ M.L. Lyra, C. Tsallis, \ \emph{Phys. Rev. Lett.} {\bf 80}, \ 53 \ (1998)\\
\noindent [25] \ C. Anteneodo, C. Tsallis, \ \emph{Phys.Rev. Lett.} {\bf 80}, \ 5313 \ (1998)\\
\noindent [26] \ U. Tirnakli, C. Tsallis, M.L. Lyra, \ \emph{Europ. Phys. Jour. B} {\bf 11}, \ 309 \ (1999)\\
\noindent [27] \ V. Latora, M. Baranger, A. Rapisarda, C. Tsallis, \ \emph{Phys. Lett. A} {\bf 273}, \ 97 \ (2000)\\
\noindent [28] \ E.P. Borges, C. Tsallis, G.F.J. Ananos, P.M.C. de Oliveira, \ \emph{Phys. Rev. Lett.} {\bf 89},\\
                              \hspace*{8mm}  25 \ (2002)\\ 
\noindent [29] \ F. Baldovin, A. Robledo, \ \emph{Europhys. Lett.} {\bf 60}, \ 518 \ (2002)\\
\noindent [30] \ F. Baldovin, A. Robledo, \ \emph{Phys. Rev. E} {\bf 66}, \ R045104 \ (2002)\\ 
\noindent [31] \ F. Baldovin, C. Tsallis, B. Schultze, \ \emph{Physica A} {\bf 320}, \ 184 \ (2003)\\
\noindent [32] \ G.F.J. Ananos, C. Tsallis, \ \emph{Phys. Rev. Lett.} {\bf 93}, \ 020601 \ (2004)\\   
\noindent [33] \ E.P. Borges, U. Tirnakli, \ \emph{Physica A} {\bf 340}, \  227 \ (2004)\\
\noindent [34] \ G.F.J. Ananos, F. Baldovin, C. Tsallis, \ \emph{Eur. Phys. Jour. B} {\bf 46}, \ 409 \ (2005)\\
\noindent [35] \ A. Celikoglu, U. Tirnakli, \ \emph{Physica A} {\bf 372}, \ 238 \ (2006)\\
\noindent [36] \ U. Tirnakli, C. Tsallis, \ \emph{Phys. Rev. E} {\bf 73}, \ 037201 \ (2006)\\
\noindent [37] \ L.G. Moyano, C. Tsallis, M. Gell-Mann, \ \emph{Europhys. Lett.} {\bf 73}, \ 813 \ (2006)\\
\noindent [38] \ J.A. Marsh, M.A. Fuentes, L.G. Moyano, C. Tsallis, \ \emph{Physica A} {\bf 372}, \ 183 \ (2006)\\
\noindent [39] \ C. Tsallis, \ \emph{Physica A} {\bf 365}, \ 7 \ (2006)\\
\noindent [40] \ H.J. Hilhorst, G. Schrehr, \ \emph{J. Stat. Mech.} {\bf P06003} \ (2007)\\
\noindent [41] \ A. Rodriguez, V. Schwammle, C. Tsallis, \ \emph{J. Stat. Mech} {\bf P09006} \ (2008)\\
\noindent [42]  \ N. Kalogeropoulos, \  \emph{Physica A} {\bf 391}, \ 3435 \ (2012)\\ 
\noindent [43]  \ J. Cheeger, D.G. Ebin, \ \emph{Comparison Theorems in Riemannian Geometry}, \\
                     \hspace*{8mm} AMS Chelsea \  (1975)\\ 
\noindent [44]  \ T. Sakai, \ \emph{Riemannian Geometry}, \ Amer. Math. Soc. \ (1996)\\
\noindent [45] \  M.R. Bridson, A. Haefliger, \emph{Metric Spaces of Non-Positive Curvature}, Springer (1999)\\
\noindent [46] \ L. Nivanen, A. Le Mehaut\'{e}, Q.A. Wang, \ \emph{Rep. Math. Phys.} {\bf 52}, \ 437 \ (2003)\\ 
\noindent [47] \ E.P. Borges, \ \emph{Physica A} {\bf 340}, \  95 \  (2004)\\
\noindent [48]  \ T.C. Petit Lob\~{a}o, P.G.S. Cardoso, S.T.R. Pinho, E.P. Borges, \  \emph{Braz. J. Phys.} {\bf 39},\\
                          \hspace*{8mm}   402 \  (2009) \\          
\noindent [49]  \ N. Kalogeropoulos, \ \emph{Physica A} {\bf 391}, \ 1120 \ (2012) \\
\noindent [50] \ E. Ghys, P. de la Harpe, (Eds). \  \emph{Sur les Groupes Hyperboliques d'apr\`{e}s Mikhael \\
                             \hspace*{8mm} Gromov}, \  Birkh\"{a}user \ (1990) \\
\noindent [51] \ A. Katok, B. Hasselblatt, \ \emph{Introduction to the Modern Theory of Dynamical \\ 
                    \hspace*{8mm} Systems}, \  Cambridge Univ. Press \ (1995)\\                   
\noindent [52] \  L. Barreira, Y. Pesin, \ \emph{Dynamics of Systems with Nonzero Lyapunov Exponents}, \\  
                    \hspace*{8mm} Cambridge Univ. Press \ (2007)\\       
\noindent [53] \  S.M. Gersten, \ \emph{Geom. Funct. Anal.} {\bf 4}, \  37 \ (1994)\\

\end{document}